\documentclass[12pt]{article}
\usepackage{amsmath}
\usepackage[margin=2.0cm]{geometry}
\usepackage{epsfig}
\usepackage{cite}
\usepackage{amsfonts}
\usepackage{fancyhdr}
\usepackage{setspace}
\usepackage{enumitem}
\usepackage{graphicx}
\usepackage{subcaption}
\usepackage[font={sf,sl}]{caption}
\usepackage{bm}
\usepackage{cancel}
\usepackage{wasysym}
\usepackage{amssymb}
\usepackage{wrapfig}
\usepackage{xcolor}
\usepackage{textcomp}
\usepackage{arydshln}
\usepackage{tikz}
\usepackage{authblk}
\usepackage{titlesec}
\usepackage{hyperref}
\usepackage[export]{adjustbox}
\usepackage[page,toc,titletoc,title]{appendix}
\usetikzlibrary{arrows}
\newcommand{\authorname}{Vineet Dawara} 
\titleformat{\section}
{\normalfont\bfseries\scshape}{\thesection}{1em}{}

\titleformat{\subsection}
{\normalfont\bfseries\scshape}{\thesubsection}{1em}{}

\titleformat{\subsubsection}
{\normalfont\slshape}{\thesubsubsection}{1em}{}

\titleformat{\title}{\normalfont\bfseries}{\thesection}{1em}{}

\title{\vspace{-3em} \large \bfseries Design of a low-velocity impact framework for evaluating space-grade materials}
\author[1]{\normalsize \authorname}
\author[1]{\normalsize Ashok Bajantri}
\author[1]{\normalsize Harish Singh Dhami}
\author[2]{\normalsize SVS Narayana Murty}
\author[1]{\normalsize Koushik Viswanathan\thanks{koushik@iisc.ac.in}}
\affil[1]{Department of Mechanical Engineering, Indian Institute of Science, Bangalore}
\affil[2]{Indian Space Research Organization, India}

\date{\vspace{-1em}\normalsize \textsc{\today}}
\numberwithin{equation}{section}
\numberwithin{table}{section}
\begin{document}

\maketitle
\thispagestyle{plain}
\vspace{-1em}
\hrulefill
\begin{abstract}
Material deformation and failure under impact loading is a subject of active investigation in space science and often requires very specialized equipment for testing. In this work, we present the design, operational analysis and application of a low-velocity ($\sim 100$ m/s) projectile impact framework for evaluating the deformation and failure of space-grade materials. The system is designed to be modular and easily adaptable to various test geometries, while enabling accurate quantitative evaluation of plastic flow. Using coupled numerical methods and experimental techniques, we first establish an operating procedure for the system. Following this, its performance in two complementary impact configurations is demonstrated using numerical and experimental analysis. In the first, a Taylor impact test is performed for predicting the deformed shape of a cylindrical projectile impinging on a rigid substrate. In the second, deformation of a plate struck by a rigid projectile is evaluated. In both cases, physics-based models are used to interpret the resulting fields. We present a discussion of how the system may be used both for material property estimation (e.g., dynamic yield strength) as well as for failure evaluation (e.g., perforation and fracture) in the same projectile impact configuration.
\end{abstract}
\paragraph*{Keywords -} Gas gun, Large strain deformation, projectile impact, low-velocity impact
	
\section{Introduction}\label{Introduction}

Failure of metallic materials under projectile impact is a subject of active investigation in several sub-domains of space science. For instance, mitigating satellite failure due to space debris requires materials and structures that are resilient enough to withstand high strain-rate deformation of $\sim 10^3 - 10^5$ s$^{-1}$ \cite{akahoshi2008influence, krag20171, liou2006risks}. While extensive work has been carried out on hypervelocity ($>1$ km/s) impact events in the context of space debris in low-earth orbit \cite{silnikov2018numerical, poniaev2017hypervelocity}, relatively lesser work has focused on low-velocity ($10-100$ m/s) impact events. However, these events are equally important to understand---as an example, Whipple shields can be significantly less effective at low velocity impact due to complete projectile penetration and lack of fragmentation \cite{yasaka2000low}. While hypervelocity impact is common in low earth orbit applications, low-velocity events are expected to be more likely for deep-space missions. 

The performance of a complex structure is commonly obtained by piecing together behaviour of representative samples under individual unit impact or deformation events \cite{nayan2015microstructure, shockey2007shear}. These events can help drive the design of new materials and/or structures with engineered internal features \cite{selivanov2021using, dudziak2015harpoon}. Quantifying impact-induced failure is also vital for evaluating the performance of more commonplace materials such as structural steels and glass sandwich panels \cite{chen2017numerical, kim2011forming} that are used for spacecraft and launch vehicle applications.

There is hence a fundamental need for a well-controlled impact testing system for materials performance evaluation, especially in the context of space applications. Unfortunately, while many impact test systems exist in research laboratories across the world (e.g., see Ref.~\cite{yasaka2000low}), with several research results being continuously published, such a system is difficult to reproduce from scratch, due to the paucity of available design information, operating procedures and calibration data. This manuscript attempts to fill that gap by outlining the design and performance of a projectile impact test-bed. We provide the basic design, detailed performance evaluation, as well as demonstrations of the use of this setup in two common impact configurations \cite{ruan2017high}. In the first---the so-called Taylor impact test---a deformable projectile is struck against a comparatively rigid target plate. This configuration is most often used for material property evaluation---the yield strength of the projectile material may be evaluated from its deformed shape \cite{taylor1948use}. The second test, termed plate impact, is a complementary configuration, in which a comparatively rigid projectile is impacted against a deforming target plate. Here, the focus is on evaluating failure mechanisms in the target material. In either case, specific microstructural mechanisms such as adiabatic shear banding and crack bifurcation can be correlated with impact parameters and kinematically measured deformation fields \cite{rogers1979adiabatic, yl1992adiabatic, meyers1994dynamic, viswanathan2020shear, zhou1996dynamically}. These mechanisms are, in general, not amenable to direct real-time measurements, necessitating the use of techniques such as \lq quick stop\rq\ projectiles to infer temporal evolution \cite{wingrove1973influence}. Results of these experiments often suggest the use of structural and geometrical changes to either accentuate or mitigate the occurrence of plastic failure in the projectile/target or both \cite{sankar2016effect, faye2015mechanics, rittel2002shear}.

The primary kinematic/geometric parameters in a typical projectile impact test in either configuration described above are the projectile velocity, size and shape. The desired projectile velocity is achieved using a gas gun, wherein an inert gas (such as He, Ar or N$_2$) is compressed to a suitable initial pressure and allowed to expand rapidly in order to accelerate the projectile. The result is projectile exit velocities that can range from 10-250 m/s \cite{swift2005light}. The accelerated projectile is retained within a barrel and strikes a target plate placed a fixed distance away. One of several impact models may then be employed to evaluate failure in either the projectile or the target, as a function of impact parameters \cite{anderson1988ballistic, zhang2004modified, backman1978mechanics}. 


While the setup described and analyzed in this manuscript is not unlike common frameworks described before, we present a complete design-to-performance description using a combination of quantitative mechanics models and experimental observations for both impact configurations. The primary purpose of the framework is to evaluate failure mechanisms in space-grade materials, that are of specific interest to the Indian Space Research Organization. The design does not focus on maximizing projectile velocity; rather, the emphasis is on achieving versatility in conducting different projectile-target impact experiments. The experimental configuration is also easily adaptable for realizing larger target sizes and a wider variety of target shapes to simulate actual parts. We envisage this manuscript as being the first in a series of works that explore a systematic method for designing lightweight structural materials for space applications.

The manuscript is organized as follows. The basic design and realization details of the impact framework are discussed in Sec.~\ref{sec:gasgundesign}. The main results are presented in three parts. Firstly, coupled numerical simulations and experimental velocity measurements are used to evaluate performance envelopes in Sec.~\ref{sec:gasgunperformance}. Following this, we present investigations of the two impact configurations described above---Taylor impact (Sec.~\ref{sec:TaylorImpact}) and plate deformation (Sec.~\ref{sec:PlateTest}). In each of these tests, experimental results are accompanied by a presentation of mechanics-based models to quantitatively understand the final deformation. In all three parts, the findings are followed by a short exposition on how the corresponding results may be used in a practical setting. Finally, a broad overall discussion is presented in Sec.~\ref{sec:conclusion}, along with some concluding remarks.


\section{Experimental configuration}	
\label{sec:gasgundesign}
\begin{figure}[h!]
	\centering
	\begin{subfigure}[b]{0.7\columnwidth}
		\includegraphics[width=\linewidth]{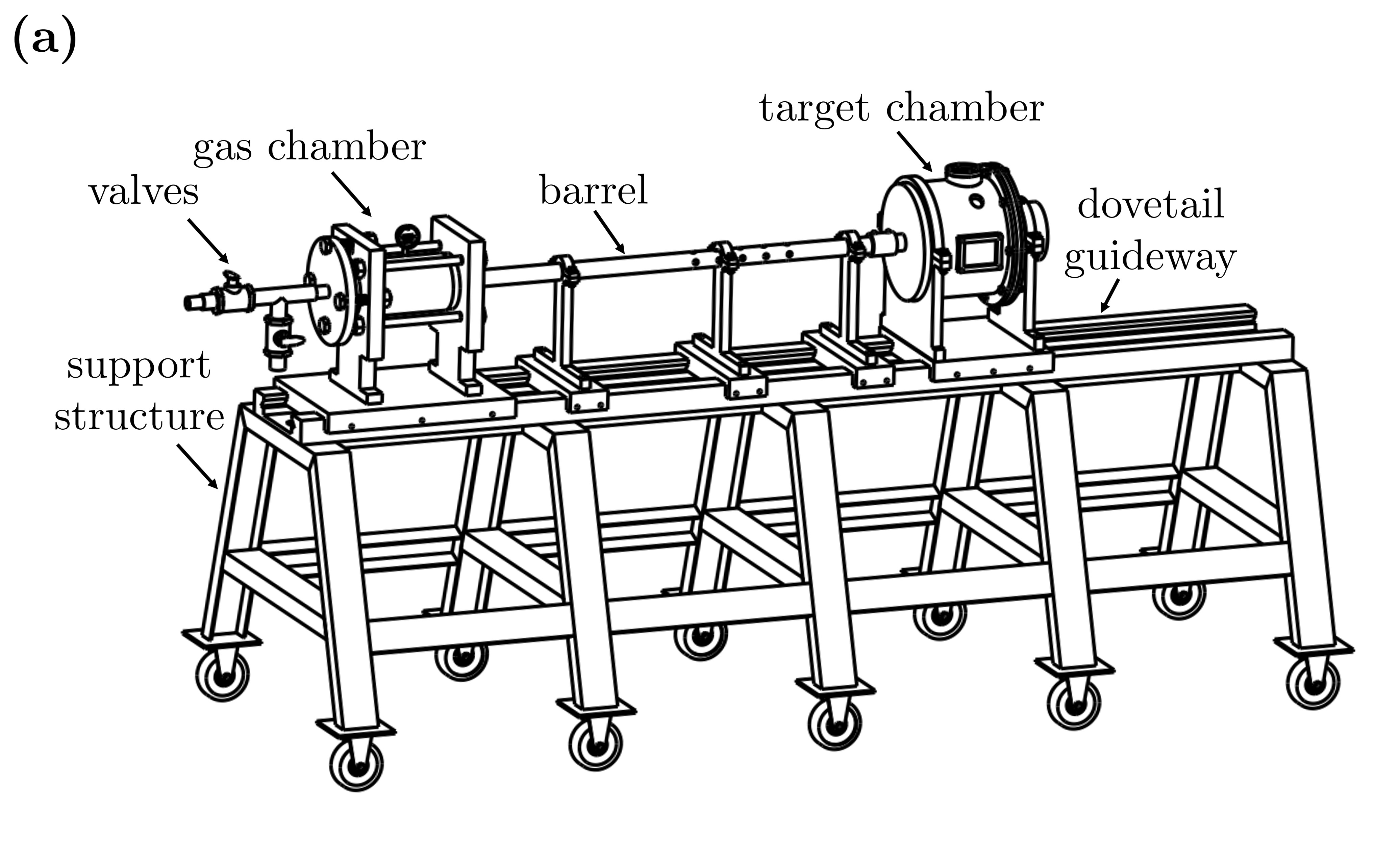}
	\end{subfigure}
   \vspace{0.5 cm}
	\begin{subfigure}[b]{\columnwidth}
		\includegraphics[width=\linewidth]{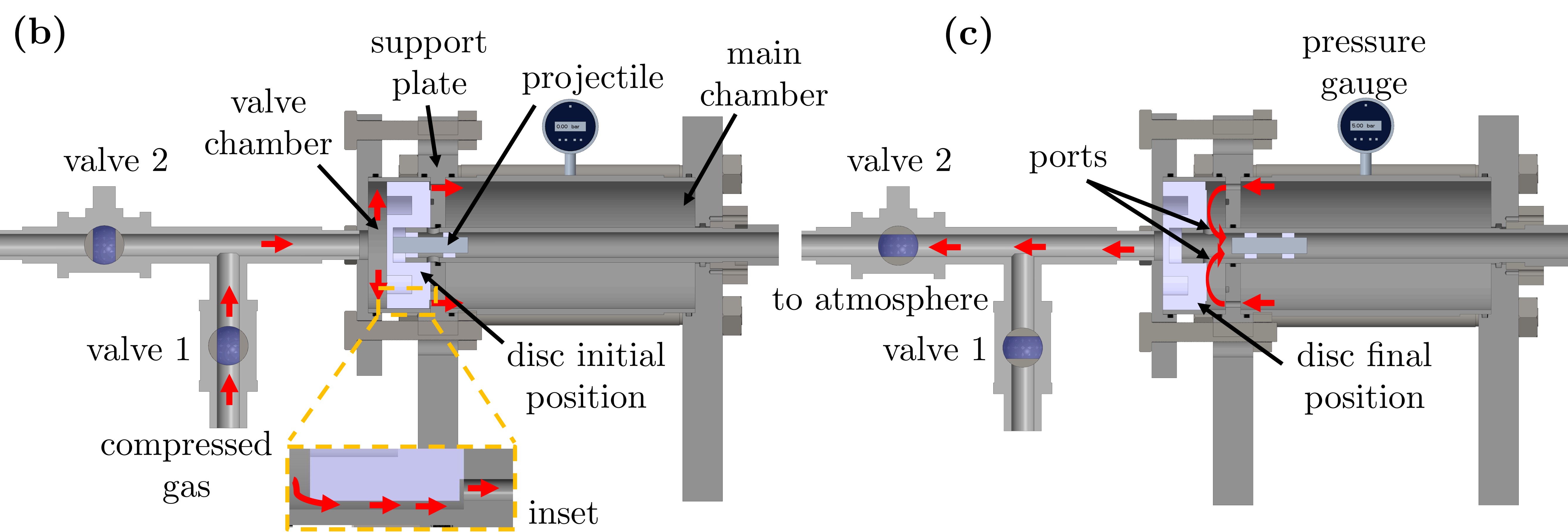}
	\end{subfigure}
	\vspace{0.5 cm}
	\caption{Operational drawings of the gas gun system. (a) Schematic of the complete system showing the gun, barrel and target chamber mounted on guideways for easy of disassembly. Panels (b) and (c) show the operation mechanism using the two-valve system in sequence. The projectile is at the breech end of the gun, and is fired by air escaping from the main chamber.}
	\label{fig:GG}
\end{figure}

The primary components of the impact framework are a gas gun, a barrel of fixed length and a target chamber, housing the target, see Fig.~\ref{fig:GG}(a). These are mounted on dovetail guides fixed to a supporting structure. The overall dimension of the setup is 2.5 m $\times$ 1 m $\times$ 1 m. Dovetail mounting enables easy projectile loading, while also allowing changes to individual components to be made quickly. The detailed description of individual components, as well as the triggering mechanism and velocity measurement system, are discussed below.

\subsection{Gas chambers and triggering mechanism}
\label{subsec:gaschamber}
The gas gun has a total volume of 1.8 litres and is designed to withstand a maximum pressure of 50 bar, see Fig.~\ref{fig:GG}(b). The overall design and triggering mechanism is based on that commonly used for split-Hopkinson pressure bars (see, for instance, Ref.~\cite{sharma2011split}). The gun has two internal chambers, termed valve and main chambers. A two-valve arrangement regulates gas flow from an external compressed air source (here a portable air compressor) to the gas chamber. The projectile is typically supported by coaxial Teflon rings, called sabots. One or two sabots are used, depending on the projectile length. Their use is necessary because sabots minimize friction between the projectile and interior barrel surface, while also helping align projectiles of arbitrary diameter centrally with the barrel axis. The projectile is first loaded from the target, through the barrel and close to the end connected to the gun (termed the breech end). This is done by moving the target plate on the dovetail and using a long enough tube to push the projectile through the length of the barrel. 

The actuation mechanism to accelerate the projectile is then operated as follows, see Fig.~\ref{fig:GG}(b) and (c). When valve 1 is opened and valve 2 closed, pressurized gas from the external source flows into the valve chamber. This gas then enters the main chamber via the clearance provided between the sliding disc and interior surface of the valve chamber. O-rings are provided on the sliding disc and support plate to prevent gas from flowing directly to the barrel. Once the main chamber is filled to the required pressure (measured using a pressure gauge), valve 1 is closed. The projectile placed at the breech end of the barrel is now triggered by opening valve 2. When this is done, pressure inside the valve chamber drops and high-pressure gas from the main chamber flows back, pushing the sliding disc. It enters the barrel immediately through the ports, see Fig.~\ref{fig:GG}(c), providing the required driving force for projectile motion.

This simple triggering system is inherently advantageous as it does not require any operating skill, and can be easily reconfigured for subsequent tests. Further, experiments have been performed to demonstrate that a projectile velocity of $>100$ m/s can be easily achieved by simply using compressed air, with higher velocities attainable using a light gas such as He \cite{seigel1965theory}.

\subsection{Barrel and velocity measurement system}
\label{subsec:velocitymeasurement}
\begin{figure}
  \includegraphics[width=\linewidth]{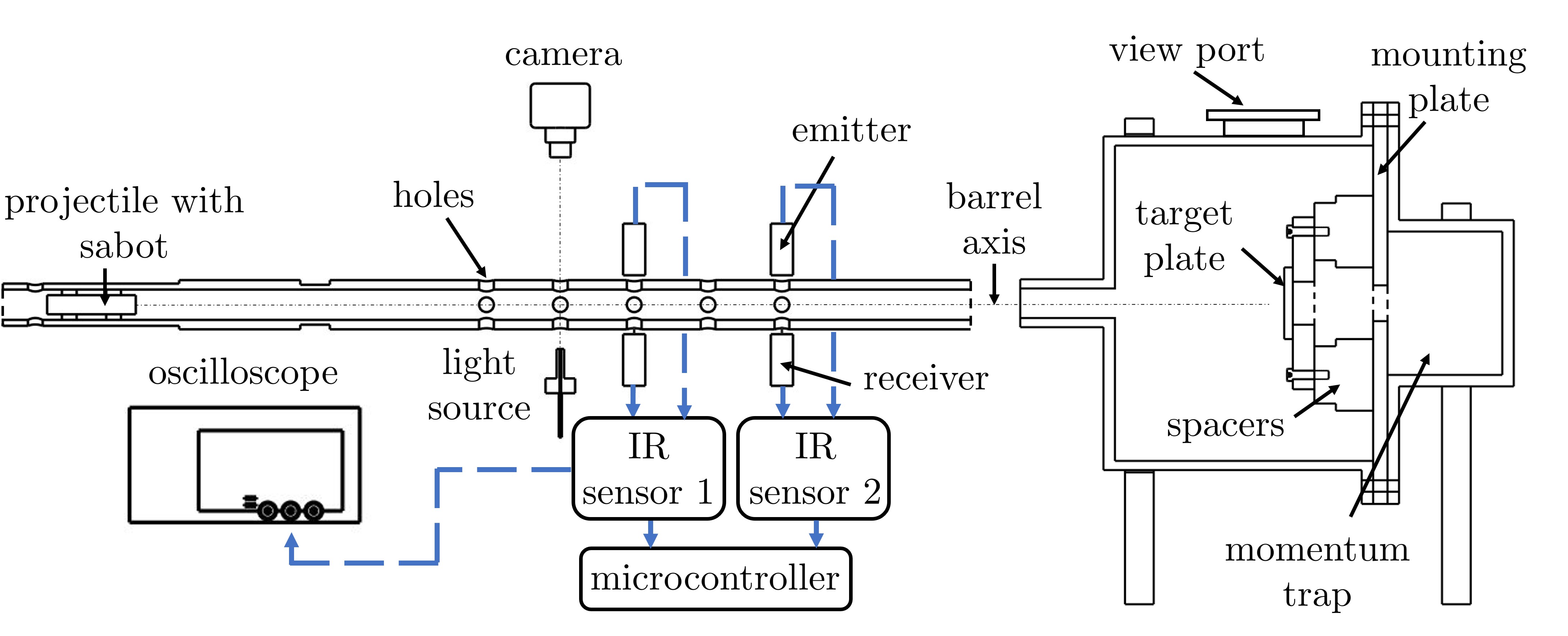}
  \caption{Measurement methods for determining projectile exit velocities from the barrel. (Left) Section near the barrel end showing vent holes and the three independent measurement systems---a camera/light source combination, IR sensors with a microcontroller and an oscilloscope for interrupt measurements. (Right) Target chamber cross section showing the plate mounting details and momentum trap.}
  \label{fig:velMeasurement}
\end{figure}

The barrel housing the projectile has an inner diameter of 21 mm and total length of 1.4 m. Five sets of four vent holes (10 mm diameter, 50 mm apart, arranged at $90^\circ$ from each other) are provided along the length of the barrel for allowing gas to escape, see Fig.~\ref{fig:velMeasurement}. If this is not done, the possibility of high gas pressure resulting in secondary impact may obscure the results. The location of the first hole, at a distance of 740 mm from the barrel breech end, is chosen to maximize projectile exit velocity, while preventing secondary impacts. The velocity measurement system consists of IR sensors and a microcontroller (Arduino Mega 2560), as shown in Fig.~\ref{fig:velMeasurement}. An oscilloscope is also used in parallel with the microcontroller for verifying the measurements. A camera and light source arrangement may also be employed across opposite lying ventholes for this purpose.

Projectile velocity within the barrel, and prior to entering the target chamber, is measured using two IR sensors that read signals from the third and fifth holes, spaced 100 mm apart, see Fig.~\ref{fig:velMeasurement}. Each IR sensor consists of an emitter and receiver placed at opposite facing vent holes on the barrel. The sensor's voltage output changes from low to high at the instant when the projectile is at the vent hole, and is recorded by the microcontroller. The projectile velocity is estimated the distance between these holes, divided by the time difference between the two crossing events. 

A second velocity measurement is performed using the voltage output of the IR sensor placed at the third hole, using a digital oscilloscope (GW Instek GDS 11028). When the projectile passes through the hole, it obstructs the receiver, and the voltage output changes from high to low and back. The width of this signal gives the time taken by the projectile to pass through the hole, and the velocity can be estimated by dividing the length of the projectile by the obtained time interval.

Finally, a third velocity measurement, for comparison, is also performed by capturing a high speed image sequence through the second vent hole using a high-speed camera (Photron AX100). The camera and light source are placed opposite each other, see Fig.~\ref{fig:velMeasurement}. Initially, the camera records a bright image that changes to dark as the projectile passes. By counting the number of dark images $n$ and noting the frame rate (FPS) set for the camera, the velocity is estimated as
\begin{align}
V_0 = \frac{L_0}{n/\rm{FPS}}
\end{align}
where $L_0$ is the length of the projectile. All three measurement methods gave identical values (within $\pm 1\%$) for the projectile velocity and are hence reported interchangably in the rest of the manuscript.

\subsection{Target chamber}  
The fully enclosed target chamber consists of a circular mounting fixture with holes for mounting the target plate, see Fig.~\ref{fig:velMeasurement}. This circular fixture is easily replaceable for mounting different sizes of targets, varying from 50-120 mm (square) and 150-200 mm (circular) dimensions. Appropriate spacer lengths can be provided to change the distance between the barrel exit and the target plate, to accommodate load cells for force measurement. Three viewing windows---two parallel to each other and the other perpendicular, are provided for imaging purposes. A momentum trap, made either of wood or silicone rubber is placed at the end to arrest projectile motion in case of complete target perforation.

\section{Results I: Gas gun performance evaluation }
\label{sec:gasgunperformance}
The basic requirement of a single impact experiment is that a projectile of mass $m$ be accelerated to a pre-determined velocity $V_0$. Since the gas gun is pressure-controlled, we require a scheme for evaluating the projectile exit velocity as a function of the filled chamber pressure $p_0$ and $m$. This is obtained using numerical 1D gas dynamics calculations and results in so-called velocity-pressure curves. These numerical results are then compared with experimental measurements to provide an operation scheme for the gun.

\subsection{Numerical 1D gas dynamics calculations}
\label{subsec:1Dgasdynamics}
As discussed in Sec.~\ref{subsec:gaschamber}, as gas exits the main chamber, it causes the projectile to move forward. This is essentially effected via compression and rarefaction waves that develop inside the main chamber, causing pressure variations behind the projectile\cite{seigel1965theory}. Since the pressure behind the projectile determines the force moving it forward, these waves influence its velocity during motion inside the barrel. Given the finite length of the main gas chamber, reflected waves are soon set up and travel forward towards the projectile, further lowering the pressure behind it. An analysis of these pressure fluctuations as well as simulations of projectile velocity are performed using the description in Ref.~\cite{seigel1965theory}. Only the final results are presented in this section. 

\begin{figure}
	\centering
	\includegraphics[width=\columnwidth]{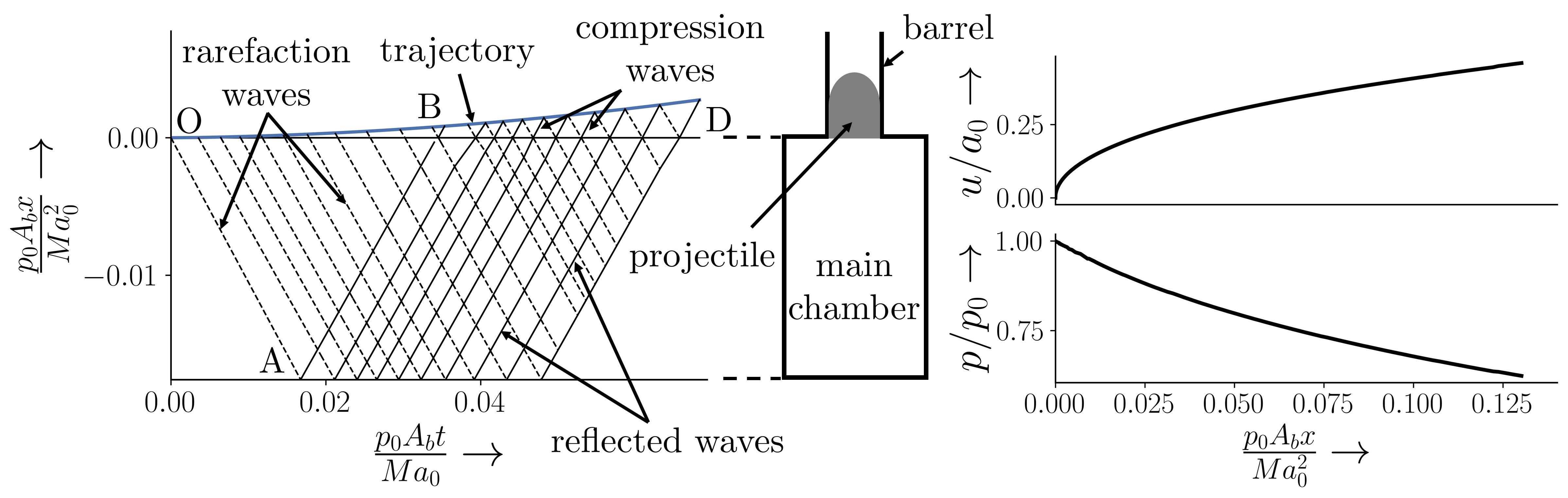}
	\caption{Gas dynamic calculations for dynamic gas pressure and exit velocity of projectile. (Left) A space-time diagram showing characteristic curves for rarefaction waves behind the projectile. Corresponding sections of the main gas chamber and breech end of the barrel are shown in the middle schematic. Solid (dashed) lines correspond to $u+a$ ($u-a$) characteristics (Right) Variation of pressure and velocity with distance along the barrel. Projectile mass $M = 10$ g at initial pressure $p_0 = 5$ bar, $A_b$ is the cross-section area of the barrel, and intial speed of sound $a_0 = 435$ m/s.}
	\label{fig:CharacteristicsSolution}
\end{figure}
Figure~\ref{fig:CharacteristicsSolution}(left) shows the gas dynamics calculation for one such situation when the initial gas pressure is set to $5$ bar, the projectile mass is $10$ g, $A_b$ is the cross-sectional area of the barrel, and $a_0 = 435$ m/s is the initial speed of sound. The gas is assumed to be ideal with a specific heat ratio of $1.4$. The horizontal (time) and vertical (distance along barrel) axes are presented in non-dimensional form using $p_0, M, a_0$ and $A_b$. The transition (line OD) from the main chamber to the barrel is assumed to be sudden in this analysis. The solid and dashed lines in this figure are the two characteristic lines corresponding to isentropic one-dimensional gas flow, given by

\begin{equation}
\frac{D(u\pm \sigma)}{Dt} = 0\;\;\text{along}\;\;u\pm a\;\;\text{characteristic lines} \label{eq: characteristice}
\end{equation}
where $u$ is the gas velocity, $a$ is the speed of sound in gas (in general, not constant), $\sigma = \int dp/(a\rho)$ is the Riemann function, with $p$ and $\rho$ the pressure and density of the gas, respectively. The $u-a$ ($u+a$) lines, shown by dashed (solid) lines, correspond to disturbances moving away from (towards) the projectile. Only a few characteristic lines are drawn, along with the chamber and barrel schematic (middle). When the projectile moves, it sends corresponding disturbance (rarefaction waves) towards the main chamber, see Fig.~\ref{fig:CharacteristicsSolution}(left). As these waves arrive at the transition between barrel and main chamber, they are partly reflected towards the projectile (compression waves) and partly transmitted (rarefaction waves) to the main chamber, where they are further reflected (reflected waves) on arriving at the other end. In this figure, OAB consists of only rarefaction waves, with no effect of reflected waves, and is termed the simple wave region. When reflected waves return to the transition, they again undergo partial reflection/transmission away from/towards the projectile.

The characteristic gas equation, Eq.~\ref{eq: characteristice}, is solved using a step-by-step numerical calculation, simultaneously with Newton's second law for tracking the projectile's trajectory. The result is a gas velocity $u$ and pressure $p$ immediately behind the projectile as it traverses the length of the barrel, see Fig.~\ref{fig:CharacteristicsSolution}(right). Here $u$ and $p$ are non-dimensionalized using the speed of sound in static air $a_0$ and initial gas pressure $p_0$. The horizontal distance $x$ along the barrel is again non-dimensionalized by a characteristic length scale $M a_0^2/p_0 A_b$. It is clear from the $u$ vs. $x$ graph that the projectile velocity tends to saturate after about $x \sim 0.1 M a_0^2/(A_b p_0)$. At this length, the gauge pressure behind the projectile also reduces significantly, resulting in negligible force. For a given $p_0$, the projectile exit velocity $V_0$ is hence $u$ evaluated at $x$ equal to the distance to the first vent hole.

\subsection{Experimental measurement: velocity-pressure curve}
\label{Sec_exp_vel}

The results of the 1D calculations provide the projectile exit velocity $V_0$ as a function of filling pressure $p_0$ for various $m$. To calibrate this prediction for the system described in Sec.~\ref{sec:gasgundesign}, experimental data was obtained for various projectiles, see Table~\ref{Tab:ProjectileDetails}. Friction between the sabot and the (stainless steel) barrel was neglected since the corresponding friction coefficient is very small (0.05--0.08, see Ref.~\cite{constantinou1987frictional}). Two sabots, totalling one-third of the projectile length, were used with all projectiles, except for the smallest one (mass 10 g). For this case, owing to length constraints, a single sabot with one-third length was used instead.

\begin{table}[h!]
	\renewcommand{\arraystretch}{2}
	\begin{center}
		\textbf{\caption{Blunt cylindrical projectile dimensions for velocity measurements}\label{Tab:ProjectileDetails}}
		\begin{tabular}{c c c c c}
			\textbf{Combined (projectile+sabot) mass (g)} & 10 & 24 & 35 & 50\\
			\hline
			\hline
			\textbf{Material} & Al & Al & SS & SS\\
		
			\textbf{Initial diameter $D_0$(\rm{mm})} & $10.0$ & $14.0$ & $13.0$&$13.7$\\
		
		\textbf{Initial length $L_0$ (\rm{mm})} & $26.1$ & $50.0$ & $30.5$&$44.0$\\
			
			\hline
		\end{tabular}
	\end{center}
\end{table}	

\begin{figure}
	\centering
	\includegraphics[width=0.85\columnwidth]{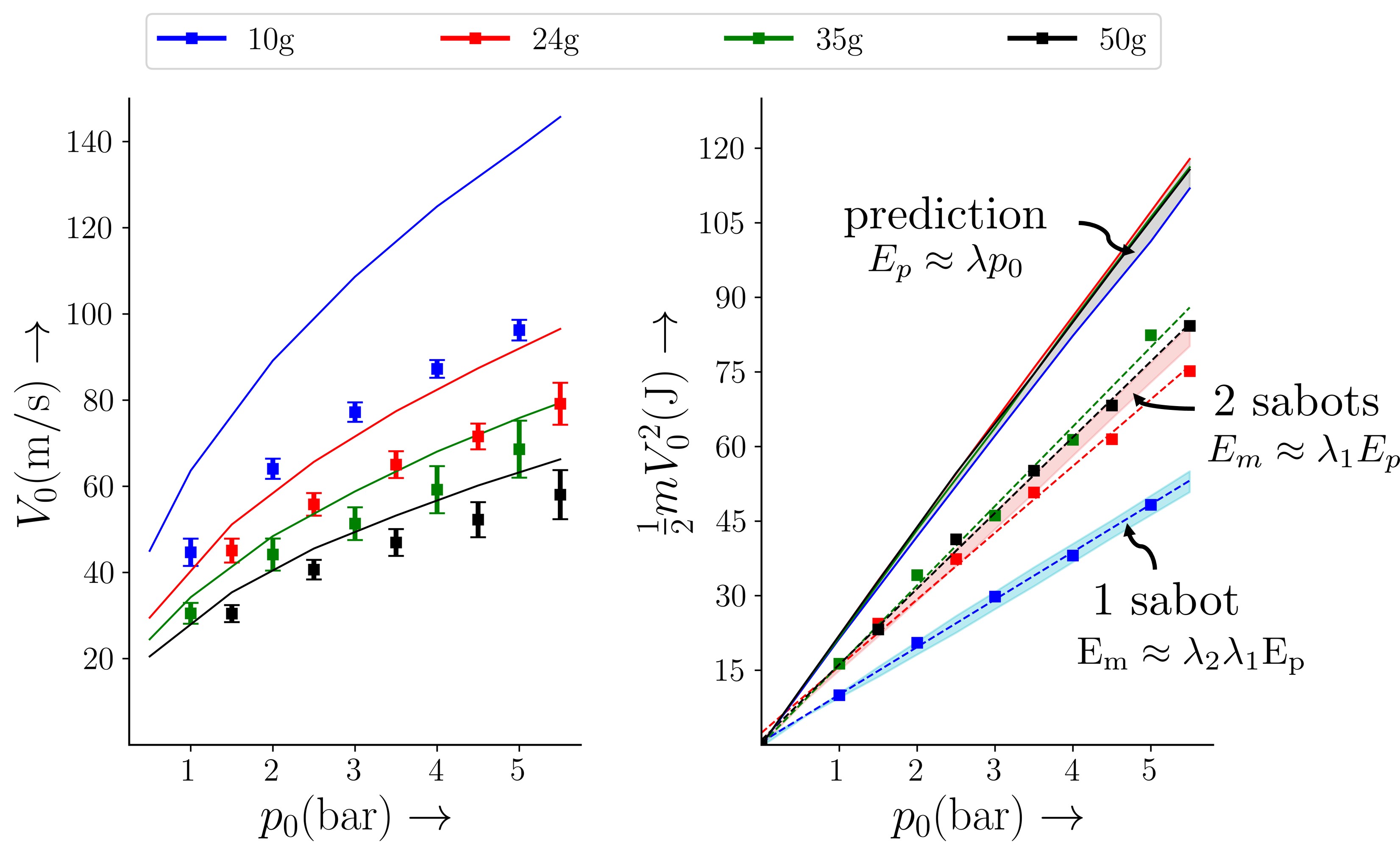}
	\caption{An operating velocity-pressure curve (left) and kinetic energy-pressure curve (right) for the gas gun setup presented in this work. Theoretical lines (solid lines) and experimental measurements (square markers) are shown superimposed. The dashed lines are fits to the experimental data, and shaded regions indicates theoretical bands as described in the text.}
	\label{fig:CalibrationCurve}
\end{figure}

Before presenting the experimental measurements, the following distinctions between the theoretical model and the actual configuration in Sec.~\ref{sec:gasgundesign} must be considered. Firstly, since significant gas escape occurs when the projectile crosses the first set of vent holes, the 1D gas analysis computation is only performed until the first hole. Secondly, in the analysis, the main chamber is assumed to be connected to the barrel by a sudden transition region, while the actual design uses a more gradual transition. Hence, the gas chamber contains $10\%$ lesser gas volume than considered in the model. As a first approximation, this volume correction is incorporated by decreasing predicted projectile kinetic energy by $10\%$. The final predicted relation between the projectile exit velocity $V_0$ and the initial gas pressure $p_0$ is summarized in Fig.~\ref{fig:CalibrationCurve} (left), for various values of (combined projectile $+$ sabot) mass $m$. This graph, termed a velocity-pressure curve, provides the initial filling pressure $p_0$ required for accelerating a combined mass $m$ to the required $V_0$.

Experimental validation of this theoretical curve is performed using a total of 8 to 10 velocity readings for a given $m$, using simultaneous independent velocity measurements (\emph{cf.} Sec.~\ref{subsec:velocitymeasurement}). The mean velocity (square marker) is plotted over the predicted curves in Fig.~\ref{fig:CalibrationCurve} (left). Error bars represent standard deviation of the measurements. From the data in this figure, it is clear that projectiles can be propelled to $V_0 \sim 100$ m/s at $p_0 = 5.5$ bar. While compressed air was used for these results, the velocity can be nearly doubled if a light gas, such as He, is used instead \cite{hutchings1975simple}.

A systematic discrepancy is clear in Fig.~\ref{fig:CalibrationCurve}(left) between the predicted and measured $V_0$ curves, primarily because the numerical model assumes an idealized one-dimensional geometry, whereas flow from the main chamber to the barrel is actually two-dimensional, as described in Sec.~\ref{subsec:gaschamber}. To account for this systematic deviation, we note the following primary observations from the plot of kinetic energy ($=1/2 mV_0^2$) of the projectile as a function of initial pressure $p_0$, see Fig.~\ref{fig:CalibrationCurve}(right). Firstly, all the theoretical predictions (solid lines) fall within a very narrow band (shaded grey region, labelled \lq prediction\rq ) approximated by a single straight line ($E_p \approx \lambda p_0$) with slope $\lambda = 20.9$ J/bar. Secondly, the measured values are also found to lie on fitted straight lines (dashed lines) for each projectile mass. These best-fit lines have slope lesser than $\lambda$, indicating that the energy loss increases with $p_0$. As mentioned earlier, except for the $10$ g projectiles, the slope of the other $V_0-p_0$ experimental lines differ only marginally; thus, the theoretical band can be scaled by factor $\lambda_1\simeq0.71$ (shaded light orange, labeled \lq 2 sabots\rq ) to coincide with the measurements. It should be recognized that this factor $\lambda_1$ accounts for the energy losses caused by two-dimensional flow from the main chamber to barrel, and backflow from the main chamber to the atmosphere via the valve chamber. Thirdly, as aforementioned, all the projectiles are supported by two sabot structures, except for the $10$ g projectile, which has only one. Moreover, with a single sabot, larger energy loss occurs  due to gas flow past the clearance between the sabot and the inner barrel surface. This gas leak was also observed when measuring velocity using high-speed imaging. To account for this, the theoretical band is further scaled by factor $\lambda_2\simeq 0.64$ (shaded light cyan region, labeled \lq 1 sabot\rq ) to match experimental values for the $10$ g, single-sabot projectile.

\subsection{Implications}
The operating regime for carrying out impact experiments using a given projectile can be selected based on Fig.~\ref{fig:CalibrationCurve}. For a given projectile of mass $m$ and required exit velocity $V_0$, the required initial operating pressure is obtained from the abscissa of the corresponding curve in this figure. 

The pneumatic triggering mechanism used here allows projectile velocities of $\sim$100 m/s with just compressed air, resulting in much higher strain rates than in conventional configurations such as Izod-Charpy or drop-weight tests\cite{ruan2017high}. This velocity range is suitable for evaluating deformation at strain rates of $10^3 - 10^5$/s. The present design does not emphasize optimizing velocity, considering the complexity involved in achieving higher velocities, such as the use of multiple stages, but instead focuses on conducting versatile laboratory-scale low-velocity impact experiments. However, as evident from the 1D gas dynamics analysis in Sec.~\ref{subsec:1Dgasdynamics}, the maximum projectile velocity in the current gas gun setup can be further increased by increasing the initial $p_0$ or by using helium \cite{hutchings1975simple}. It is also worth mentioning that such analyses are usually integrated with the design phase of the gun in order to estimate the various dimensions of the gas chamber and barrel for the desired velocity range.

In summary,  for a given initial pressure $p_0$ and projectile mass $m$, the theoretical kinetic energy can be estimated as $E_p= \lambda p_0$, then the output kinetic energy of the projectile in our setup is obtained as either $E_m= \lambda_1 E_p$ if two sabots are used, or $E_m= \lambda_2\lambda_1 E_p$ if single sabot is used.

\section{Results II: Taylor Impact test}
\label{sec:TaylorImpact}
Having established a method to determine the operating parameters, we now proceed to discuss two complementary impact tests. The first is the Taylor impact test where the projectile is deformable and the target plate is comparatively rigid. The converse test is discussed in Sec.~\ref{sec:PlateTest}.   

The Taylor impact test uses a deformable cylindrical projectile to strike a stationary plate. Compressive elastic and plastic waves start propagating towards the rear end of the projectile immediately after impact. Elastic waves, being much faster, reach the rear end first and are reflected as tensile waves, which then undergo subsequent reflection from compressive waves arising from the elastic-plastic boundary. This cyclic process results in projectile deceleration until a finite plastic region is developed at the impact end, see schematic in Fig.~\ref{fig:TaylorSchematic}. We now present theoretical analysis of this process, followed by experimental evaluation using our setup. 

\subsection{Theoretical analysis}
In his pioneering work, Taylor proposed a one-dimensional simplified model\cite{taylor1948use} to estimate the dynamic yield strength in terms of density, impact velocity, and initial/final lengths of the projectile. Several modifications to this model have been proposed subsequently, to incorporate more complex constitutive material properties and for matching with the experimental observations \cite{hawkyard1969theory, jones1987equation, jones1997engineering}. However, most of these models are one-dimensional and rely on measurements of the final dimensions of the impacted specimen \cite{jones1998elementary}.


To predict the final projectile geometry, we use an axisymmetric model in which the plastic zone geometry is approximated as the frustum of a cone \cite{chakraborty2015axisymmetric}. This model considers incompressible plastic deformation that causes bulging at the impact end and incorporates spatio-temporal variations of strain and strain rate, thus also predicting deformation history. Given the constitutive equation of the material, the deformation of the impacting projectile can be obtained by solving a set of ordinary differential equations (ODEs) in time, derived purely based on physical arguments. We use this formulation to predict the influence of the impact velocity $V_0$ on the mechanics of projectile deformation. The operating parameters required to determine the impact velocity are to be estimated from the velocity-pressure relationship curve shown (\emph{cf.} Fig.~\ref{fig:CalibrationCurve}).
\begin{figure}[h!]
	\centering
	\includegraphics[width=0.5\columnwidth]{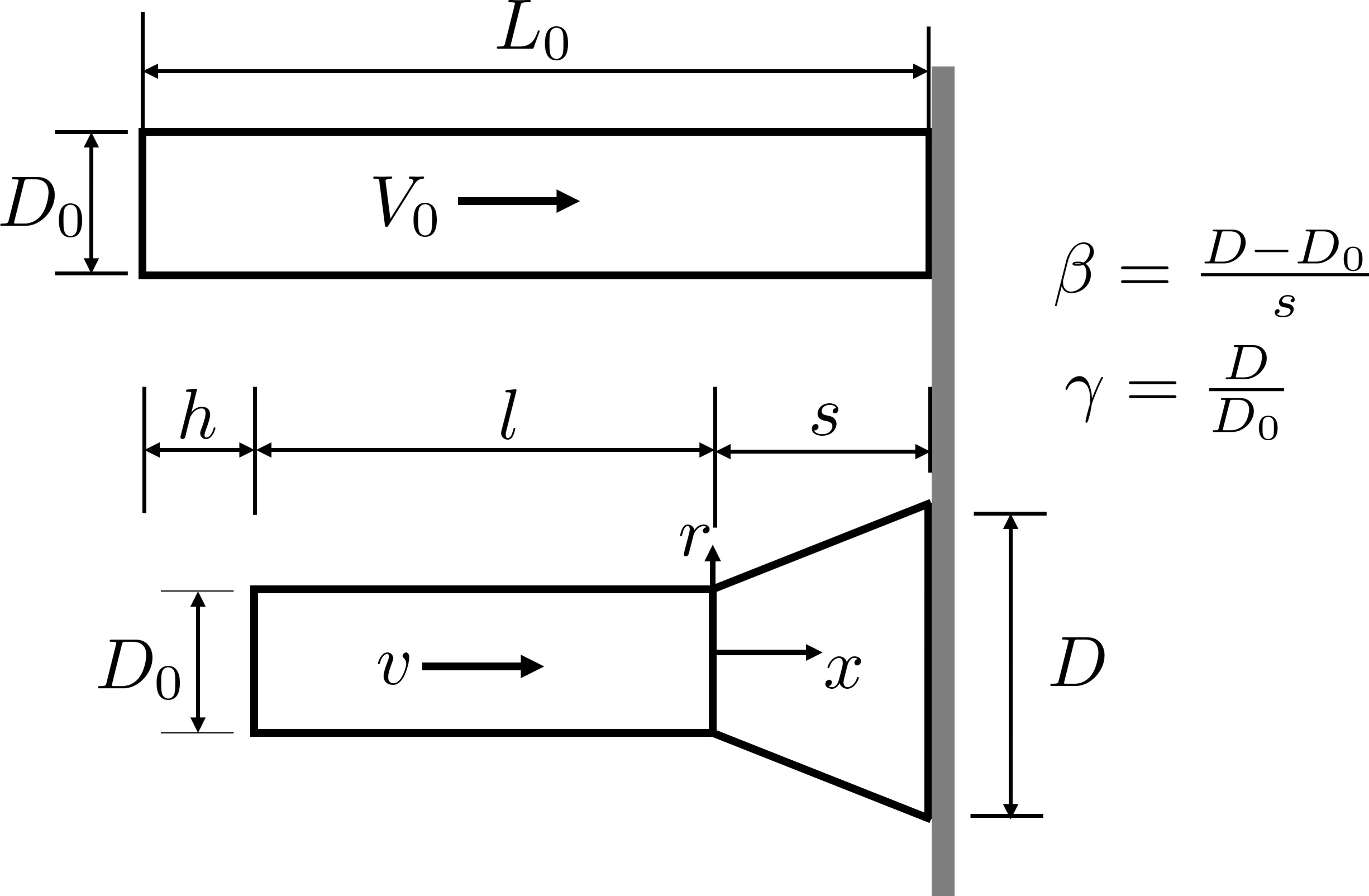}
	\caption{Schematic of Taylor impact showing the meaning of various symbols. The definitions of quantities $\beta$ and $\gamma$ are also displayed.}
	\label{fig:TaylorSchematic}
\end{figure}	

We have made two minor modifications to the model presented in Ref.~\cite{chakraborty2015axisymmetric}. Firstly, the effective plastic strain and strain rates are determined by averaging the strain rate components along the elastic-plastic boundary since the unyielded part of the projectile is subjected to stress waves generated at this boundary. The corresponding expressions for strain rate components are (\emph{cf.} Ref.~\cite{chakraborty2015axisymmetric}):
\begin{align}
	\dot{\epsilon}_{xx} &= -\frac{\beta^2 vx}{D_0(\gamma -1 - \log\gamma)(D_0  + \beta x)}\notag\\
	\dot{\epsilon}_{rr} = \dot{\epsilon}_{\theta\theta}&= \frac{\beta^2 vx}{2D_0(\gamma -1 - \log\gamma)(D_0  + \beta x)} \notag\\
	\dot{\epsilon}_{xr} &= -\frac{\beta^2 vr}{2(\gamma -1 - \log\gamma)(D_0  + \beta x)^2}\notag
\end{align}
Here, the initial diameter, length, and impact velocity of the projectile are denoted by $D_0$, $L_0$, and $V_0$, respectively. During impact, $D$ and $v$ represent the bulge end diameter and projectile velocity, see schematic in Fig.~\ref{fig:TaylorSchematic}. Also depicted here is the axial shortening $h$, undeformed length $l$, and plastic zone size $s$ that occur during impact, as well as the dimesionless ratios $\beta$ and $\gamma$. The coordinate system is located at the center of the elastic-plastic boundary (see bottom row, Fig.~\ref{fig:TaylorSchematic}). At this boundary ($x = 0$), the average strain rates are obtained by integrating over the cross-sectional area,
\begin{align}
	\dot{\bar{\epsilon}}_{xr} &= \frac{4}{\pi D_0^2}\int_{0}^{2\pi}\int_{0}^{D_0/2}\dot{\epsilon}_{xr}\Big|_{x=0}rdrd\theta = \frac{\beta^2v}{6D_0(\gamma-1-\log\gamma)}\notag
\end{align}
and the other components are identically zero ($\dot{\bar{\epsilon}}_{xx} = \dot{\bar{\epsilon}}_{rr} = \dot{\bar{\epsilon}}_{\theta\theta} = 0$). The effective plastic strain rate can now be obtained by using the von-Mises criterion:
\begin{align}
		\dot{\bar{\epsilon}}_{p} = \sqrt{\frac{2}{3}\dot{\epsilon}_{ii}} = \frac{\beta^2v}{3\sqrt{3}D_0(\gamma-1-\log\gamma)}\label{eq: epr}
\end{align}
The effective plastic strain can be determined by time integration of this equation. Additionally, the yield stress ($\sigma_y$) is determined assuming a Johnson-Cook constitutive relation, and ignoring thermal effects
\begin{align}
	\sigma_y = \left[A + B\bar{\epsilon}_p^n\right]\left[1 + C\ln\left(\frac{\dot{\bar{\epsilon}}_p}{\dot{\epsilon}_0}\right)\right]\label{eq: JC}
\end{align}
where, $A, B, n, C$ and $\dot{\epsilon}_0$ are material constants. Combining these relations, the following set of coupled non-linear ODEs are obtained
\begin{align}
	\frac{dl}{dt}  &= -c_p\\
	\frac{dh}{dt} &= v\\
	\frac{dv}{dt} &= - \frac{c_e(c_e + c_p)}{El}\\
	\frac{dD}{dt} &= \frac{\beta v(\gamma -1)}{2(\gamma - 1 - \ln\gamma)}
\end{align}
along with the expressions \eqref{eq: epr} for strain rate and the constitutive equation \eqref{eq: JC}. The symbols $E$, $c_e$, and $c_p$ denote the Young's modulus, the elastic wave speed, $c_e = \sqrt{E/\rho}$, and the plastic wave speed $c_p = \sqrt{\sigma_y/\rho}$, respectively. As pointed out in Ref.~\cite{chakraborty2015axisymmetric}, numerical instability prevails at the beginning of time integration due to the presence of $s$ in the denominator of the expression for $\beta = (D-D_0)/s$. Time integration is hence performed by setting constant values of $\beta$ and $\gamma$. 

The simulation steps are as follows---we begin with an initial guess for  $\beta$ and $\gamma$ for the first iteration and obtain the final deformed geometry. The diameter after bulging ($D_f$) is determined using the final undeformed length $l_f$ and plastic zone size $s_f$ from the previous iteration, using the incompressibility condition. Further, we update $\beta$ and $\gamma$ equal to the average values $\beta_{av} (= 0.5\beta_f)$ and $\gamma_{av} (= 0.5(1+\gamma_f))$, respectively for the subsequent iterations, until the tolerance for these quantities is below $10^{-3}$.
\begin{figure}[h!]
	\centering
	\includegraphics[width=0.8\columnwidth]{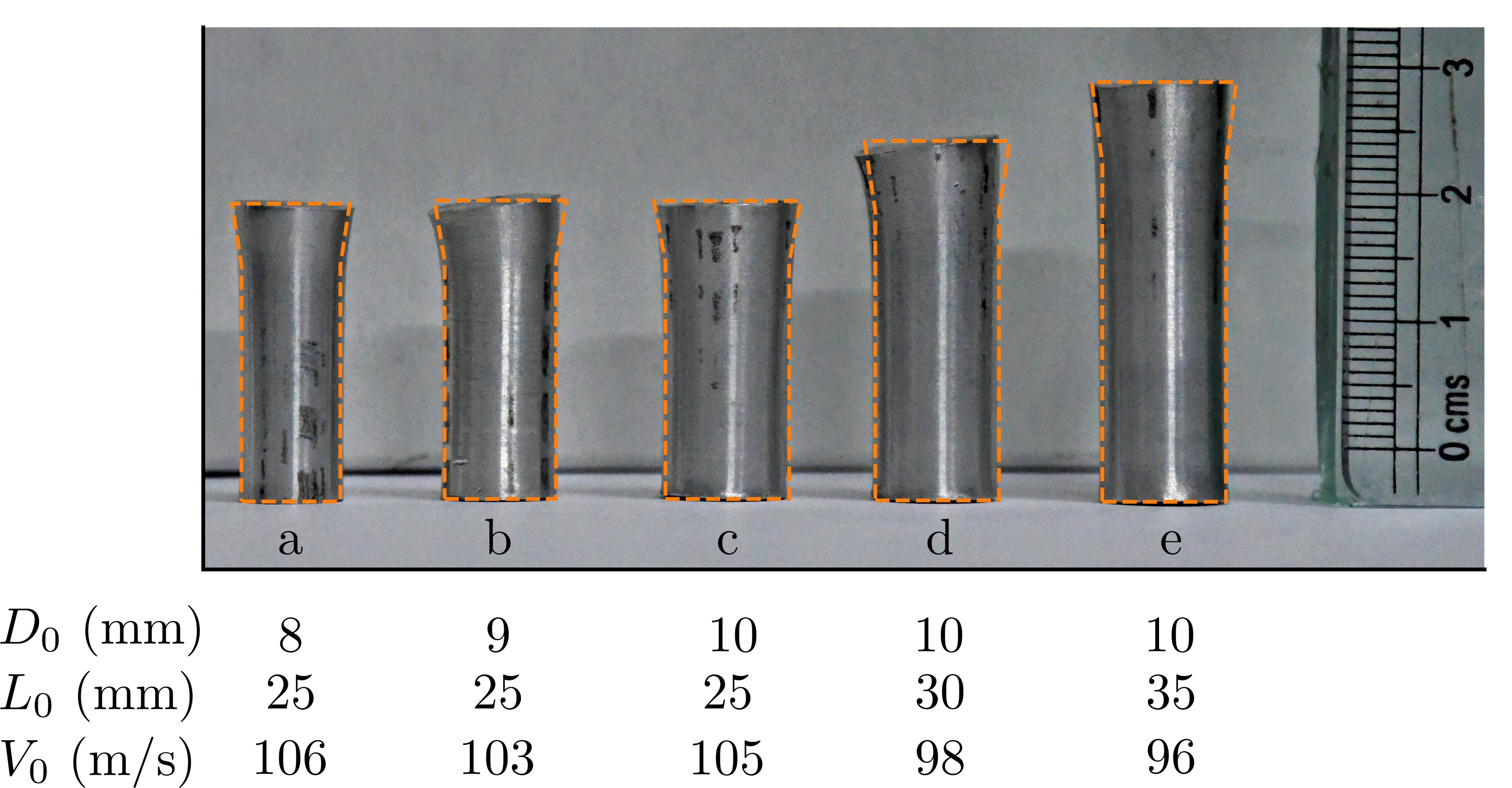}
	\caption{Taylor specimens Al 6061-T6 after impact. The initial parameters are tabulated beneath the image. The model predicted deformed geometry (in orange) is overlaid over the deformed specimens.}
	\label{fig:TaylorSamples}
\end{figure}
\begin{figure}[ht!]
	\centering
	\includegraphics[width=\columnwidth]{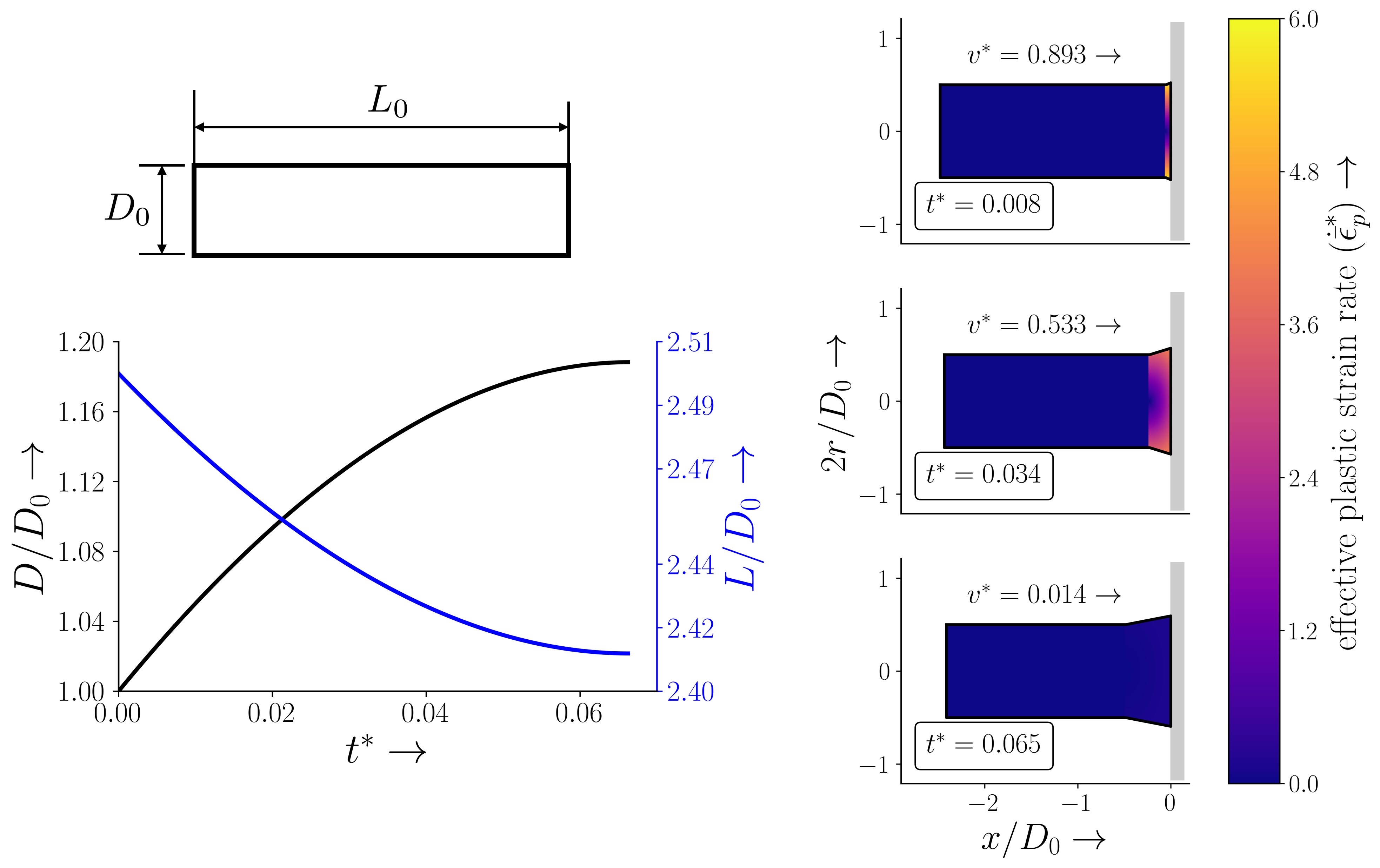}
	\caption{Numerical results for specimen 'c' ($D_0 = 10$ mm, $L_0 = 25$ mm, and $V_0 = 105$ m/s). Figure (left) shows the time history (left) of the bulge diameter $D$ and length $L$ of the projectile. Figure (right) shows the spatial effective strain field distribution ($\dot{\bar{\epsilon}}_p^*$) (right) for three distinct time instances.}
	\label{fig:Sample1}
\end{figure}

\subsection{Experimental  measurements}	
We prepared cylindrical specimens of as-received aluminum 6061 with initial diameter $D_0$ and length $L_0$ impacting an EN19 plate (20 mm thickness) with different velocities $V_0$, see Fig.~\ref{fig:TaylorSamples}. The specimens are labeled with small alphabets (\lq a\rq\ to \lq e\rq ), and their initial parameters are tabulated in the figure. The sabots used for the test consists of one Teflon ring of length 2/3 times the projectile length, placed at the free end. The combined sabot plus projectile mass for all projectiles is in the range of 12-15 g, and the corresponding operating pressure is 7 bar. Using scaling factors for single sabots (see Sec. \ref{Sec_exp_vel}), the estimated velocities are again consistent with measured values during the test. The deformed specimens are shown in Fig.~\ref{fig:TaylorSamples}, along with the predicted shapes overlaid (orange dashed lines) from the model. The close agreement between the final experimentally observed shapes and the prediction is clear from this figure. For the model computation, Johnson-Cook material constants for aluminum 6061 were taken from Ref.~\cite{lesuer2001modeling} as: $A = 334$ MPa, $B = 114$ MPa, $n = 0.42$, $C = 0.002$ and $\dot{\epsilon}_0$ = 1. We also observed minor buckling in the deformed specimens, alongside the cone-shaped bulging of the plastic zone.
	
The strain rate varies from a maximum at the beginning to zero at the end of the impact. The model additionally predicts deformation history during impact, which is otherwise challenging to obtain experimentally, see Fig.~\ref{fig:Sample1}. Temporal variation of the frustum diameter $D$ and length $L$ of the projectile (corresponding to specimen \lq c\rq ) is shown in Fig.~\ref{fig:Sample1}(left). Corresponding spatial variation of the effective plastic strain rate ($\dot{\bar{\epsilon}}_p^*$) for three different time instances is shown in Fig.~\ref{fig:Sample1}(right). All quantities are reported here in non-dimensional form, time $t^* = tV_0/L_0$, velocity $v^* = v/V_0$ and strain-rate ($\dot{\bar{\epsilon}}_p^*$) is expressed in terms of $t^*$. These results clearly predict the large instantaneous strain-rate that the bulge region near the impact end is subjected to, making it particularly susceptible to premature damage. It is expected, especially given the close match in final shapes seen in Fig.~\ref{fig:TaylorSamples}, that the spatio-temporal evolution matches that in the experiments, as may perhaps be confirmed by careful high-speed \emph{in situ} imaging. This, however, is beyond the scope of the present manuscript.

\begin{figure}[h!]
	\centering
	\includegraphics[width=\columnwidth]{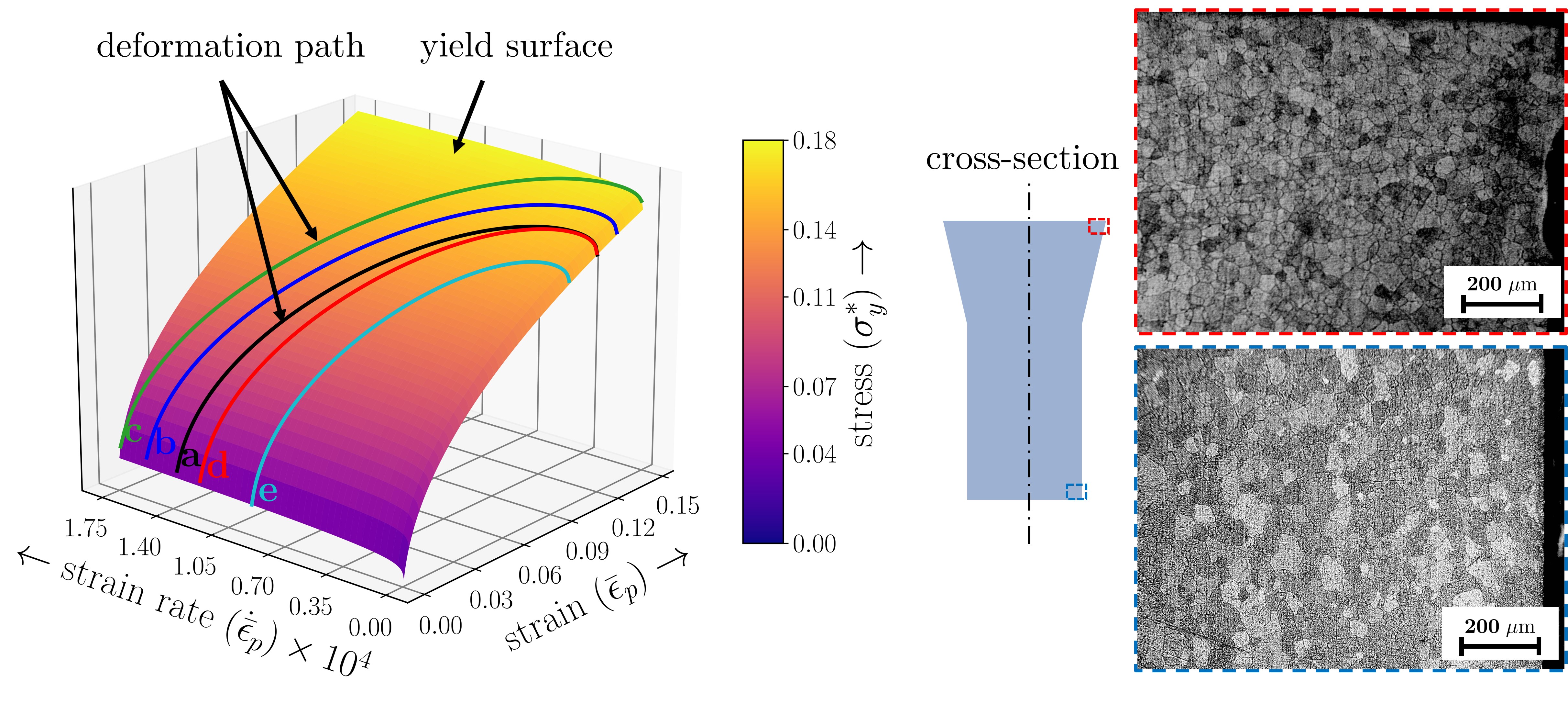}
	\caption{The deformation path (left) traced over the yield surface by all Taylor specimens  'a' to 'e'. The microstructure images (right) of a small section cut at two different locations indicated by coloured squares in the schematics on the longitudinal cross-section of specimen 'c'.}
	\label{fig:YieldSurface}
\end{figure}

Another interesting prediction of the model is the deformation path traced over the yield surface, see Fig.~\ref{fig:YieldSurface}(left). Effective plastic strain ($\bar{\epsilon}_p$), strain rate ($\dot{\bar{\epsilon}}_p$), and yield stress ($\sigma_y^*$) are all computed at the elastic-plastic boundary. Note that the non-dimensional yield stress is defined as $\sigma_y^* = (\sigma_y - A)/A$ using the Johnson-Cook parameter $A$. This deformation history is useful for semi-quantitatively predicting the final microstructure in the deformed projectile---given this information, one could easily use a multi-scale (e.g., crystal-plasticity) model to estimate grain shape/size evolution and potentially predict the onset of fracture. The varied paths of two very similar specimens \lq a\rq\ and \lq b\rq\ is also noteworthy in this figure. 

As first-attempt experimental validation, the microstructure of specimen \lq c\rq , and a schematic showing corresponding locations, is reproduced in Fig.~\ref{fig:YieldSurface}(right). The microstructure is obtained by mechanical polishing with emery papers of various grades, followed by cloth polishing using diamond paste, and then etched by submerging it for 1-3 minutes in Keller's reagent. The change in grain size is indistinguishable as the specimen is only subjected to 15\% plastic strain and 18\% more stress than the static yield stress as inferred from the deformation path of specimen \lq c\rq in Fig.~\ref{fig:YieldSurface}(left). 

\subsection{Implications}

The physics-based model described here provides a direct method to predict the deformation geometry given $m, V_0$. It also suggests two inverse problems. The first one, as originally suggested by Taylor, is to deduce the material parameters $A, B, n$ for an unknown material, from measurements of the final shape and the impact velocity $V_0$. In such a case, the inverse problem is posed in which the material parameters are perturbed in the model in such a way that the predicted deformation best fits the experimental measurements, see, for instance, Refs.~\cite{chakraborty2015axisymmetric, sen2020taylor}. Since the Taylor impact configuration is simple to set up, a number of experiments can be easily carried out using the framework presented here.

A second inverse problem is to determine the impact velocity $V_0$, given the material properties and the final deformed shape. This is of critical importance in applications where determining impact conditions are necessary as a \emph{post mortem} procedure, as for e.g., in failure of spacecraft structures. One could even go a step further and envisage predicting $V_0$ based on microstructural features in the final projectile. This will be necessary in practical situations where, perhaps part of the projectile is lost due to subseqeuent impact events and deformation history must be reconstructed from a very small portion of the final projectile. For this purpose, the relations presented in Fig.~\ref{fig:YieldSurface}, showing the deformation history of the specimen, may be exploited to obtain the initial $V_0$. This route must be taken with care, since it is quite likely that unique solutions cannot be guaranteed in general.

\section{Results III: Plate deformation experiments}
\label{sec:PlateTest}
We now discuss a configuration complementary to that of the Taylor experiment---impact of a rigid projectile against a deformable plate. Upon impact, the projectile will induce plastic deformation that may lead to fracture, and potentially, resulting in perforation \cite{backman1978mechanics}. For a given projectile-target geometry, a range of impact velocities can result in either mere initiation of plastic flow to complete perforation \cite{beynet1971plate, calder1971plastic}. We deal here only with the non-perforating case to estimate plastic deformation in the plate.

\subsection{Theoretical analysis}

Upon impact of the rigid projectile, elastic and plastic waves begin to propagate within the plate, causing transverse deflections in the direction of projectile motion. Two plastic deformation regions may be delineated---bulging where the target conforms to the projectile shape, and bending-induced dishing, which extends over a considerable distance from the impact zone.

To estimate the amount of deformation as a function of impact parameters $V_0, m$, we adopt the model proposed by Beynet and Plunkett\cite{beynet1971plate}, assuming a rigid blunt projectile and a perfectly-plastic target material with no elastic effects. The plate bends as the bulge region moves with the projectile. The plate subsequently yields due to the developed radial stress, resulting in plastic strain in the transverse direction. Assuming that the deformation is dominated by radial stress, the transverse displacement $w$ is governed by
\begin{align}
	\frac{\partial^2 w}{\partial \bar{r}^2} + \frac{1}{\bar{r}}\frac{\partial w}{\partial \bar{r}} = \frac{\partial^2 w}{\partial \bar{t}^2}\;\;\text{for}\;\;1<\bar{r}<\infty, \bar{t}\geq 0 \label{eq: waveequation}
\end{align}  	
where $\bar{r} = r/r_0$, and $\bar{t} = c_p t/r_0$ are the nondimensionalized radial distance and time, respectively. Here, $r_0$ is the projectile radius and $c_p = \sqrt{\sigma_y/\rho}$ is the plastic wave speed. Equation~\ref{eq: waveequation} describes dishing as a radial propagation of plastic strains with speed $c_p$. To determine the initial and boundary conditions for solving Eq.~\ref{eq: waveequation}, we note that the radial stress near the bulge region retards the velocity of an equivalent body consisting of projectile and bulge region, which serves as a boundary condition, and can be written as 
\begin{align}
	\alpha \frac{\partial w}{\partial \bar{r}} = \frac{\partial^2 w}{\partial \bar{t}^2},\;\;\text{at}\;\;\bar{r} = 1, \bar{t} \geq 0 \label{eq: BC}
\end{align}
with $\alpha = 2\Delta M/(M + \Delta M)$, $M$ and $\Delta M$ are masses of projectile and the bulge region, respectively. Initially, the transverse displacement is zero everywhere, and only the bulge region moves, which forms the initial conditions given by:
\begin{align}
	w(\bar{r},0) &= 0\notag\\
	\frac{\partial w(\bar{r},0)}{\partial t} &= \begin{cases}
	V_0\frac{M}{M + \Delta M}\frac{r_0}{c_p} & \bar{r} = 1\\
	0 & \bar{r} > 1
	\end{cases}	\label{eq: IC}
\end{align}
with $V_0$ the projectile velocity at the onset of impact. Equation~\eqref{eq: waveequation} is then numerically solved using a finite-difference scheme, subject to the initial (Eq.~\eqref{eq: IC}) and boundary (Eq.~\eqref{eq: BC}) conditions to obtain the final shape of the plate.

\subsection{Experimental measurements}
\label{sec:PlateExp}

\begin{figure}[h!]
	\centering
	\includegraphics[width=\columnwidth]{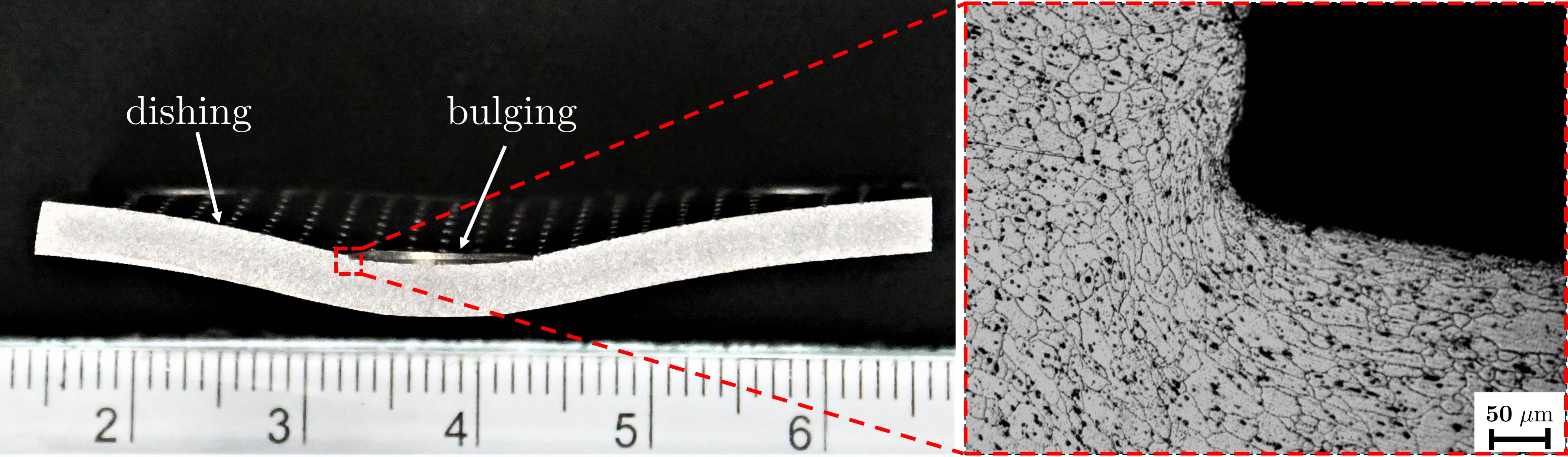}
	\caption{Deformed plate of Al 6061 (left) along with microstructure (right) of the small section near the bulge region marked in red.}
	\label{fig:PlateImpact}
\end{figure}

For obtaining experimental measurements to evaluate the predicted temporal target deformation, we impact a stainless steel projectile of diameter 10.2 mm and length 45.8 mm with $V_0 = 65$ m/s against an aluminum 6061 (as-received) target plate of dimensions 50 mm $\times$ 50 mm $\times$ 3 mm. The resulting shape is shown in Fig.~\ref{fig:PlateImpact}(left), with bulging and dishing regions demarcated. The corresponding microstructure (Fig.~\ref{fig:PlateImpact}(right)) shows clearly distinguishable grains in these two regions. The occurrence of locally compressed grains within the bulge region is evident in this figure, and this can be related to the projectile impact velocity $V_0$ using grain-size analysis.

Since the model predicts transverse displacement, we obtain equivalent experimental data by measuring the deformed surface profile of the plate. While this profile can be accurately reconstructed using laser scanning, we use a more primitive, yet easily implementable, method instead. This is done by tracing grid points embossed \emph{a priori} onto the plate's surface during preparation, see supplementary material. Using the simple fact that a static liquid surface will form a perfectly horizontal surface, we recover $x,y,z$ coordinates of points on the surface of the deformed plate. A more comprehensive description of this method is provided in the supplementary material. The final transverse displacement and developed surface strain fields in the plate can be estimated from the deformed profile, and by correlating initial and final grid locations, respectively.

\begin{figure}[h!]
	\centering
	\begin{subfigure}[b]{0.42\columnwidth}
		\includegraphics[width=\linewidth]{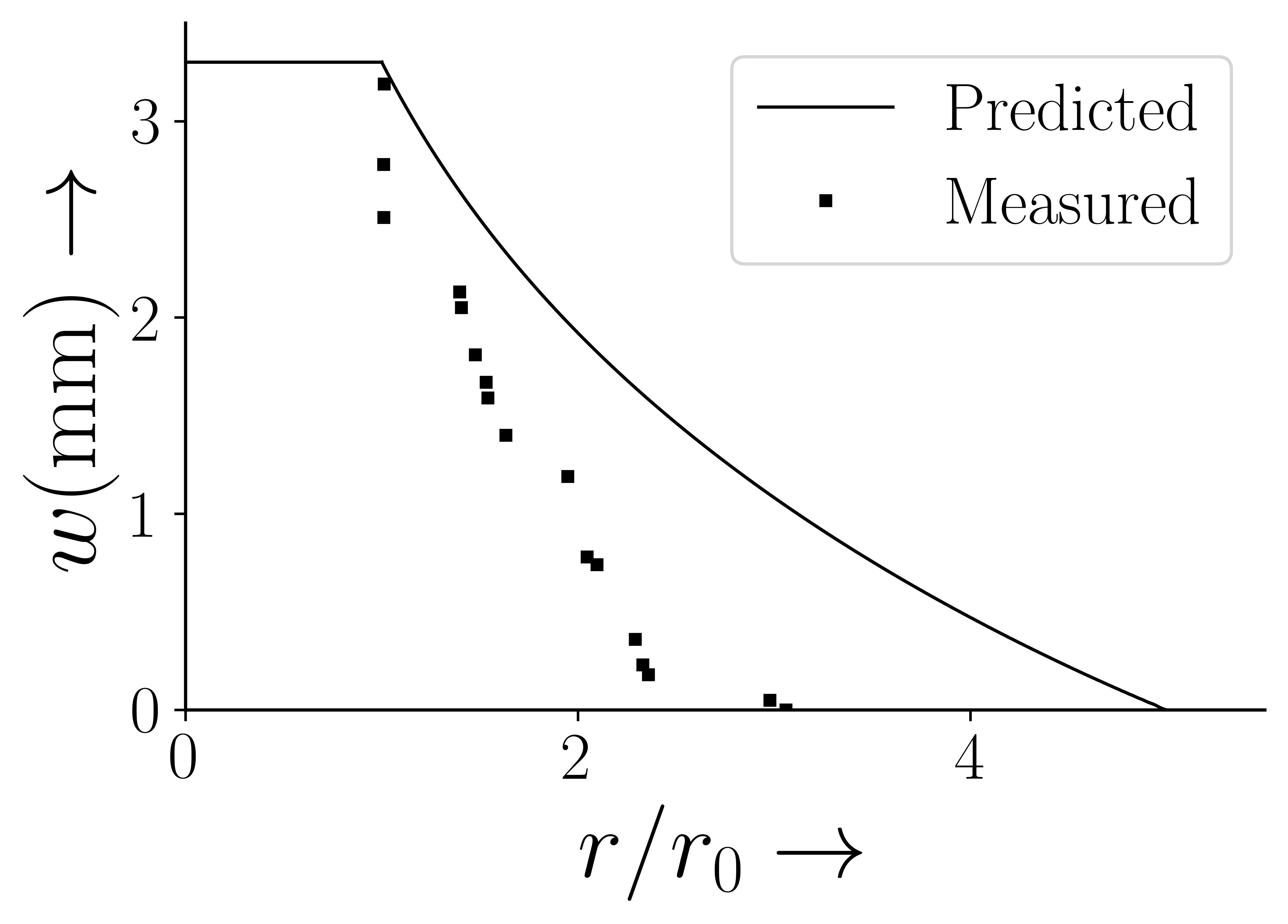}
	\end{subfigure}
	\begin{subfigure}[b]{0.55\columnwidth}
		\includegraphics[width=\linewidth]{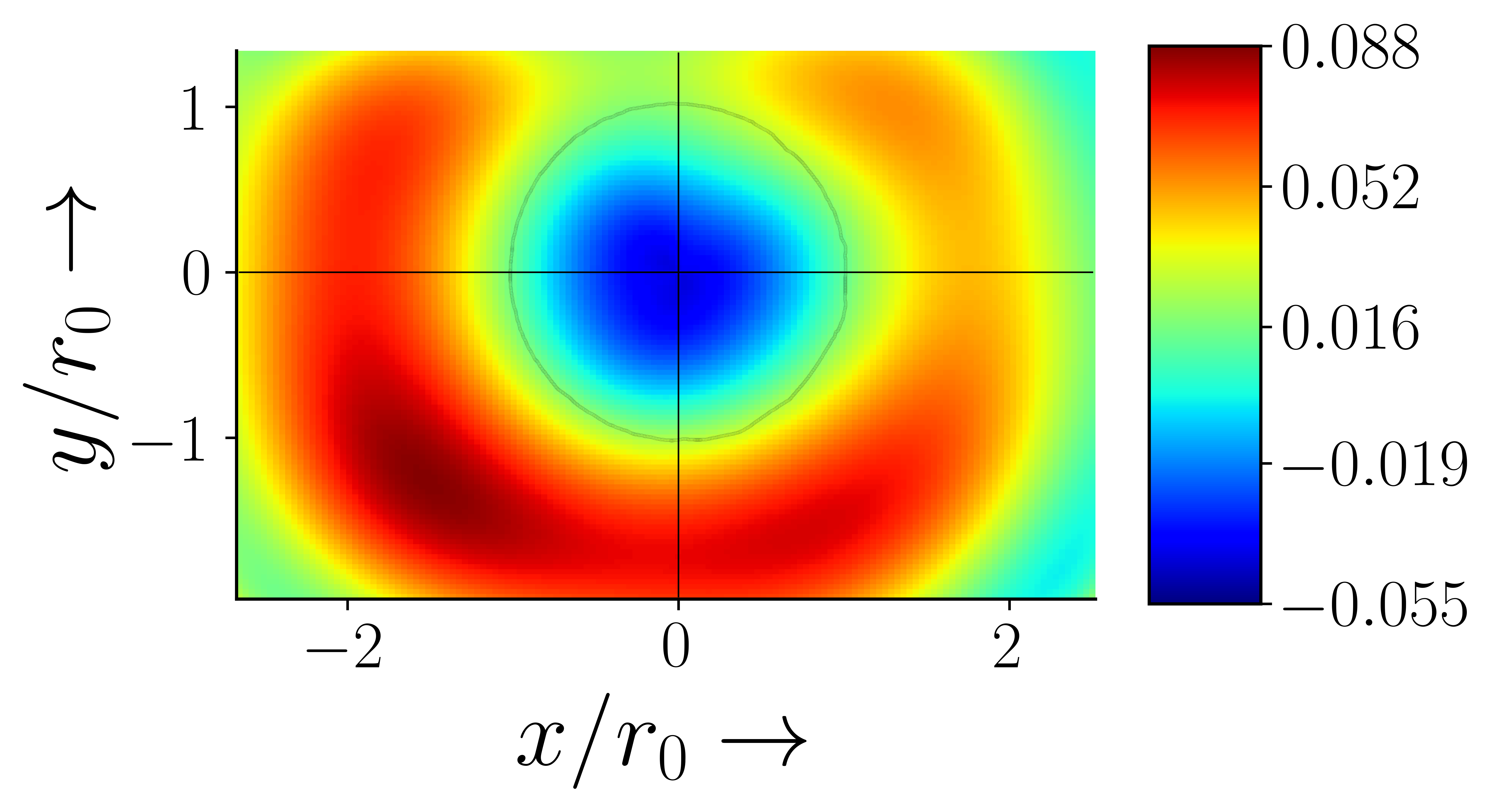}
	\end{subfigure}
	\caption{Results of the measurement using the camera-ink setup for the deformed aluminum plate. Figure (left) shows the transverse deflection $w$ as a function of radial distance $r$ from the center of the projectile impact predicted from the model compared with the experiment. The strain field component $\epsilon_{rr}$ using camera-ink setup is shown in Figure (right). }
	\label{fig:PlateSolution}
\end{figure}
	
A radial section of the obtained profile gives the deflection  $w$ of the points in the deformed plate plotted against the predicted deflection from the model, see Fig.~\ref{fig:PlateSolution}(left). For comparison, $\sigma_y = 334$ MPa and $\rho = 2700$ kg/m$^3$ were chosen as representative for Al6061. The model assumes the plate to be infinite, without accounting for edge effects from clamping. In contrast, the considered plate in the test is finite; hence, the predicted deformation is much smaller than that reconstructed from the final shape. We expect the discrepancy between the two results to reduce for larger in-plane plate dimensions.

The in-plane strain field components for a finite deformation are computed by using the displacement mapping function from the original to the deformed profile. The radial component $\epsilon_{rr}$ is only shown here, see Fig.~\ref{fig:PlateSolution}(right), estimated in the red region marked in Supplementary Figure 1. The other components $\epsilon_{\theta\theta}$ and $\epsilon_{r \theta}$ are found to be less than 10\% of the maximum value of $\epsilon_{rr}$, which \emph{a posteriori} justifies the radial symmetry assumption in the model. 

\subsection{Implications}
Despite being based on simplified theoretical arguments, the analysis presented here provides predictions of material response in typical non-perforating impact. It should be noted that any material can typically exhibit a wide variety of failure mechanisms in a plate impact test \cite{backman1978mechanics}. Some common possibilities include fracture due to initial compressive waves, spalling caused due to the reflection of compressive wave from the distal boundary of the plate, plugging resulting from highly localized shear zone formation\cite{viswanathan2020shear, rittel2008thermo, wingrove1973influence} and even fragmentation \cite{kooij2021explosive, chaudhri2015dynamic}. These failure types predominantly depend on material properties, geometrical characteristics, and impact velocity. As a result, and given its practical consequences for space applications, plate deformation has attracted significant research effort for decades \cite{prosser1999acoustic, borvik2002perforation, kaboglu2018high }.

Considering the broad variability in failure response, a predictive model relating the impact velocity with the deformation mechanics is essential in setting the test operating parameters in an experiment. As per test configuration and material heterogeneities (for instance, porous solids, composite materials \emph{etc.}), one needs to resort to more involved analyses to predict the deformation \cite{anderson2017analytical, umanzor2021penetration}.


\section{Discussion and Summary}\label{sec:conclusion}
In this work, we've described a compact laboratory-scale gas gun setup for studying low-velocity impact events on both projectiles and targets. Some noteworthy features of the configuration are that it is easy to setup and adapt to multiple test modalities (e.g., Taylor impact, perforation) and realize different final target geometries (e.g., multi-layer shields, curved structures). Our work here has focused on determining an operating scheme for the test setup, as well as calibrating theoretical velocity-pressure curves with those obtained experimentally. using only compressed air, the system was demonstrated to maximum projectile velocities of $\sim$100 m/s, making it ideal for low-velocity impact studies in a laboratory setting. 

The capabilities of the presented design have been demonstrated for two standard impact test configurations \emph{viz.} Taylor and plate impact. The former has been performed by impacting an aluminum specimen against a rigid alloyed steel plate and using an axisymmetric model for predicting deformation geometry. Close agreement between experiments and the model predictions were noted, in addition to the determination of time history for the deformation fields, which are extremely challenging to obtain experimentally. In the second impact configuration, an aluminum plate is deformed by a stainless steel projectile. A physics-based model is presented to relate the plate's transverse deflection with the impact velocity. The final displacement and strain fields after impact were measured using a visual grid mapping technique, and agrees reasonably well with the numerical prediction. While accounting for the disparities between the simplified models and the actual experiments, the numerical models help estimate the effect of impact velocity on the material deformation behaviour. The derived impact velocity will then be suitably used to estimate the operating parameters for the test by referring to the produced velocity-pressure relationship curve.

We believe that the framework described in this work is easily replicable and will be useful for groups that are exploring the use of novel structural designs for space applications, such as metamaterials or microarchitected internal features \cite{kadic20193d, fleck2010micro}, for impact energy absorption applications. We also believe that it will help elucidate the impact performance of emerging classes of metallic materials such as high-entropy alloys \cite{sonkusare2020high, george2019high} that represent an area of active research interest. The question of of how these multi-component alloys fail and how such potentially catastrophic mechanisms can be mitigated either by diverse microarchitecting, or additional alloying or a combination of both, represent interesting areas for future research for space-grade materials. We are presently investigating some of these questions and hope to communicate our results in due course.

\clearpage
\bibliography{bibfile}
\bibliographystyle{vancouver}

\end{document}